\documentclass[sigconf]{acmart}
\AtBeginDocument{%
  \providecommand\BibTeX{{%
    \normalfont B\kern-0.5em{\scshape i\kern-0.25em b}\kern-0.8em\TeX}}}

\begin{document}

\title{Black-box Adversarial ML Attack on Modulation Classification}

\author{Muhammad Usama}
\email{muhammad.usama@itu.edu.pk}
\affiliation{%
  \institution{Information Technology University, Punjab, Pakistan}
}

\author{Junaid Qadir}
\email{junaid.qadir@itu.edu.pk}
\affiliation{%
  \institution{Information Technology University, Punjab, Pakistan}
}

\author{Ala Al-Fuqaha}
\email{aalfuqaha@hbku.edu.qa}
\affiliation{%
 \institution{Hamad Bin Khalifa University, Qatar}
}


\begin{abstract}
Recently, many deep neural network (DNN) based modulation classification schemes have been proposed in the literature. We have evaluated the robustness of two famous such modulation classifiers (based on the techniques of convolutional neural networks and long short term memory) against adversarial machine learning attacks in black-box settings. We have used Carlini \& Wagner (C-W) attack for performing the adversarial attack. To the best of our knowledge, the robustness of these modulation classifiers have not been evaluated through C-W attack before. Our results clearly indicate that state-of-art deep machine learning based modulation classifiers are not robust against adversarial attacks.  
\end{abstract}




\maketitle

\section{Introduction}
Machine learning (ML) especially deep ML schemes have beaten human-level performance in many computer vision, language, and speech processing tasks which were considered impossible a decade ago. This success of ML schemes has inspired the ideas of self-driving networks \cite{feamster2017and} and knowledge defined networking \cite{mestres2017knowledge} where ML schemes are profoundly utilized to ensure automation and control of networking tasks such as dynamic resource allocation, modulation classification, network traffic classification, etc.

Despite the success of ML in different modern communication and data networking applications, there are some pitfalls in the fundamental assumptions of ML schemes which can be exploited by the adversaries to craft adversarial examples in order to compromise the ML-based system. An \textit{adversarial example} is defined as an input to the ML model specially crafted by an adversary by adding a small imperceptible perturbation to the input sample to compromise the performance of the ML model. Mathematically, an adversarial example $x^*$ can be formed by adding a typically-imperceptible perturbation $\delta$ to the legitimate test example $x$ of the deployed trained classifier $f(.)$. The perturbation $\delta$ is computed by approximating the following nonlinear optimization problem provided in equation 1 where $t$ is the targeted class in case of a targeted attack or any other wrong class is the case of untargeted attack.

\begin{equation}\label{eq1}
    x^* = x + \arg \underset{\eta{_x}}{\text{min}} \{\|\eta\|: f(x + \eta) = t\} 
\end{equation}

Adversarial examples are possible because of two major faulty assumptions in ML schemes. \textit{Firstly}, the underlying data distribution experienced during the training phase of the ML model will also be encountered in the testing phase. This data stationarity is not valid for most of the real world cases and the void created by following this assumption is exploited by the adversary for crafting the adversarial examples. \textit{Secondly}, most of the ML schemes are based on the empirical risk minimization (ERM), which is an approximation of the actual unknown probability distribution. The ERM has an associated error with it which can be exploited by the adversary to make an adversarial example.

Adversarial attacks can be classified broadly into \textit{white-box} and \textit{black-box} attacks based on the knowledge of the adversary about the deployed ML model. In a \textit{white-box attack}, it is assumed that adversary has complete knowledge (hyperparameters, test data, etc.) of the deployed model whereas in a \textit{black-box attack} no such knowledge is assumed and it is assumed that the adversary can only act as a standard user and query the system for a response.  

In this paper, we have taken modulation classification (which is an important component of modern communication and data networks) as a proxy of functional areas of cognitive self-driving networks. We have performed a black-box adversarial attack on DNN-based modulation classification to highlight the brittleness of ML schemes utilized in cognitive self-driving networks.

\section{Related Work}
% \subsection{Modulation Classification Using Deep Learning}
% There are few DL based modulation classification schemes available in the literature. O'Shea at al. \cite{o2016convolutional, o2018over} used the convolutional neural network (CNN), VGG, and ResNet for performing modulation classification, they have opted for utilizing highly cited GNU radio ML RML2016.10a dataset \cite{o2016radio} which provides 11 digital and analog modulation schemes on the SNR ranging from -20 dB to 18dB. Recently, Ramjee et al \cite{ramjee2019fast} used CNN and recurrent neural networks especially long short term memory (LSTM) for modulation classification. Similarly, Zheng et al. \cite{zheng2019fusion} used CNN with feature-based fusion for solving the modulation classification problem. DL-based classification schemes have produced very good results but they are vulnerable to adversarial examples crafted by the adversary to fool the ML-based classifier to perform incorrect classification. 
% \subsection{Adversarial Attacks on Modulation Classification}
There does not exist much literature on adversarial attacks on modulation classification. Recently, Sadeghi et al. \cite{sadeghi2018adversarial} used a variant of fast gradient sign method (FGSM) attack \cite{goodfellow2014explaining} on modulation classification on CNN-based modulation classification to highlight the threat of the adversarial examples. FGSM is an adversarial sample crafting algorithm where the adversarial perturbation is calculated by taking a gradient step in the direction of the sign of the gradient of test example. Kokalj et al. \cite{kokalj2019adversarial} also crafted the adversarial examples for modulation classification by using the FGSM perturbation generation algorithm. Most of the available results on the application of the adversarial attacks are reported by using the FGSM attack.

A shortcoming with the FGSM attack is its lack of optimality in adversarial perturbation generation as FGSM was designed to quickly craft adversarial examples irrespective of the optimality and the size of the perturbation in the test example. To overcome the lack of optimality and to highlight that optimal adversarial example for modulation classification can be crafted we have used Carlini \& Wagner (C-W) attack \cite{carlini2017towards} where the adversarial examples are crafted using the following optimization process provided in equation \ref{eq3}. 

\begin{equation} \label{eq3}
\begin{split}
    \underset{\eta}{\text{minimize}} \quad \|\eta\|{_\mathcal{P}} + c . g(x{^*}) \\
    \text{such that} \quad x{^*} \in [0, 1]{^n}
\end{split}
\end{equation}

% \begin{equation} \label{eq4}
%     g(x{^*}) = \text{max} \{0, \underset{i \neq t}{\text{max}} \mathcal{Z}(x{^*}){_i} - \mathcal{Z}(x{^*}){_t}\}
% \end{equation}

\begin{figure}[h]
\centering     %%% not \center
\centerline{\includegraphics[width=0.50\textwidth]{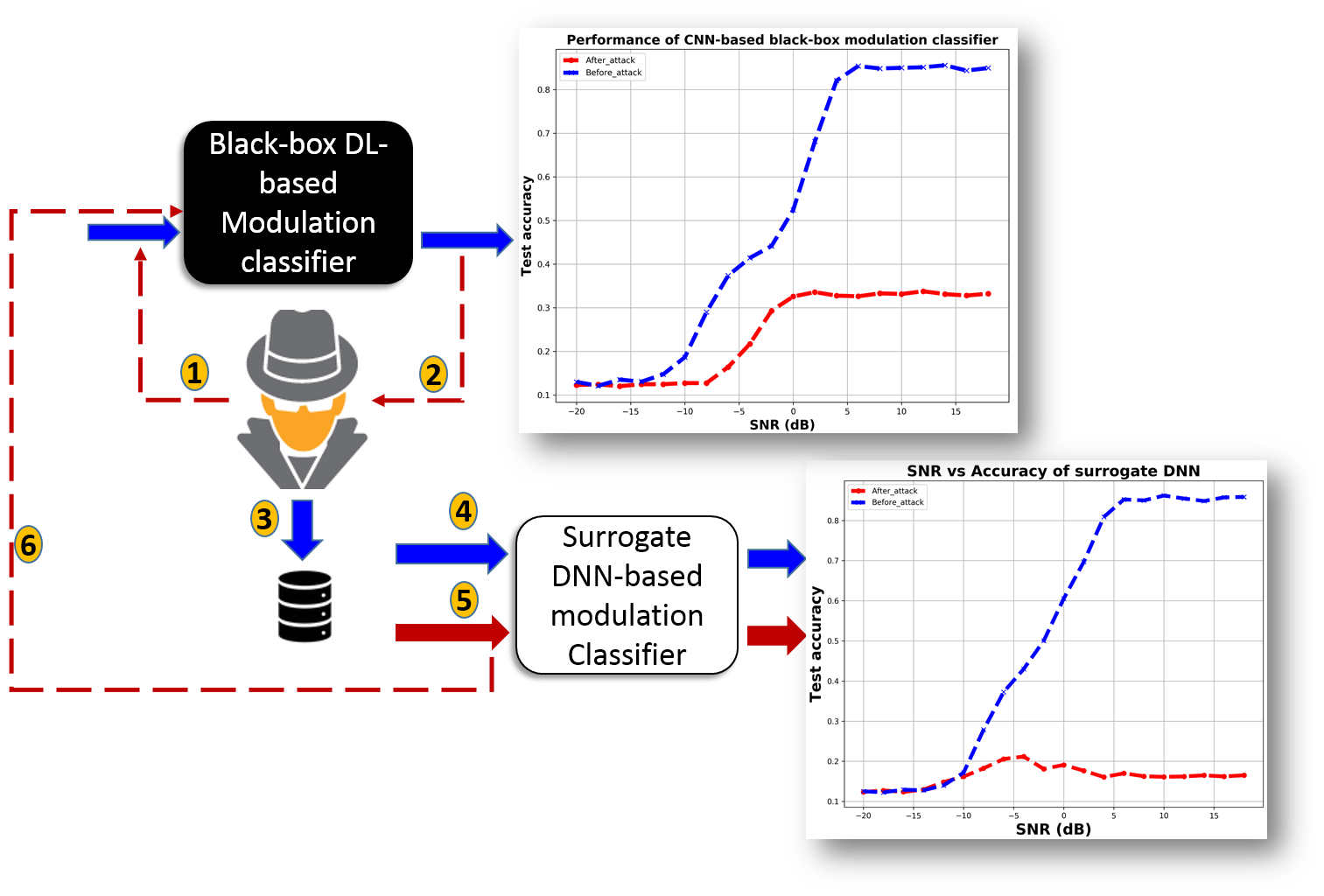}}
\caption{The step by step procedure followed for crafting black-box adversarial attack against DL-based modulation classification is depicted in the figure.}
\end{figure}

\section{Black-box Adversarial Attack Procedure}

In this section, we will provide our black-box adversarial attack procedure (illustrated in Figure 1). The steps followed are: \textbf{1)} the adversary queries the deployed modulation classifier with test examples; \textbf{2)} the deployed modulation classifier provides a labeled response to the adversary considering the adversary as a normal user; \textbf{3)} the adversary stores the query-response pair in a database (which is later used as a substitute dataset for training a surrogate DNN);  \textbf{4)} once sufficient data is collected in the adversarial database, the adversary constructs a fully connected DNN model and trains it for suitable classification performance; \textbf{5)} once the surrogate DNN is trained, the adversary launches a C-W attack on the surrogate DNN for crafting adversarial examples that compromises the performance of the surrogate DNN model; \textbf{6)} adversarial examples that compromises the performance of surrogate DNN-model are then transferred to black-box DL-based modulation classifier which according to the transferability property of adversarial examples will compromise the performance of DL-based modulation classifier.

Since we are performing this experiment in lab settings, we have opted for training two modulation classifiers based on CNN and LSTM and then considered them as black-box models. We have used highly-cited GNU radio ML RML2016.10a dataset \cite{o2016radio} which provides 11 digital and analog modulation schemes on the SNR ranging from -20 dB to 18dB. We have used only 10\% of the test examples to construct the surrogate classifier and then performed C-W attack the performance of the surrogate DNN model before and after the attack is provided in Figure 1. Once the adversarial attack on surrogate DNN is completed, we have transferred the adversarial examples that evaded the surrogate DNN to black-box modulation classifier by leveraging the transferability property of adversarial ML. The performance impact of the adversarial attack is provided in Figures 1 and 2. A clear drop in the accuracy of the modulation classifier after the adversarial attack highlights that our method of performing black-box adversarial attack has successfully compromised the performance crafted adversarial examples.

\begin{figure}[h]
\centering     %%% not \center
\centerline{\includegraphics[width=0.30 \textwidth]{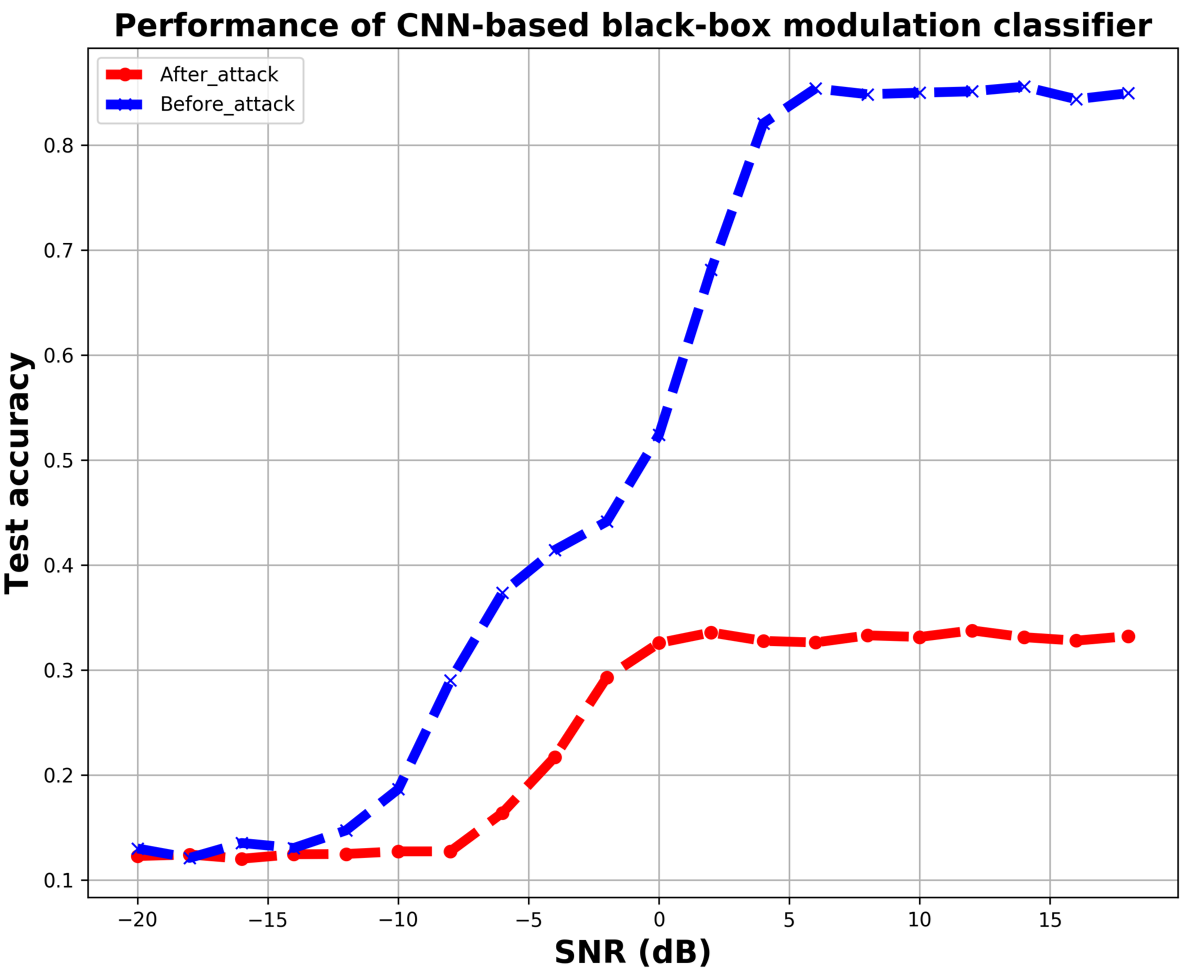}}
\caption{Performance of black-box adversarial attack on CNN-based modulation classification.}
\end{figure}

\begin{figure}[h]
\centering     %%% not \center
\centerline{\includegraphics[width=0.30 \textwidth]{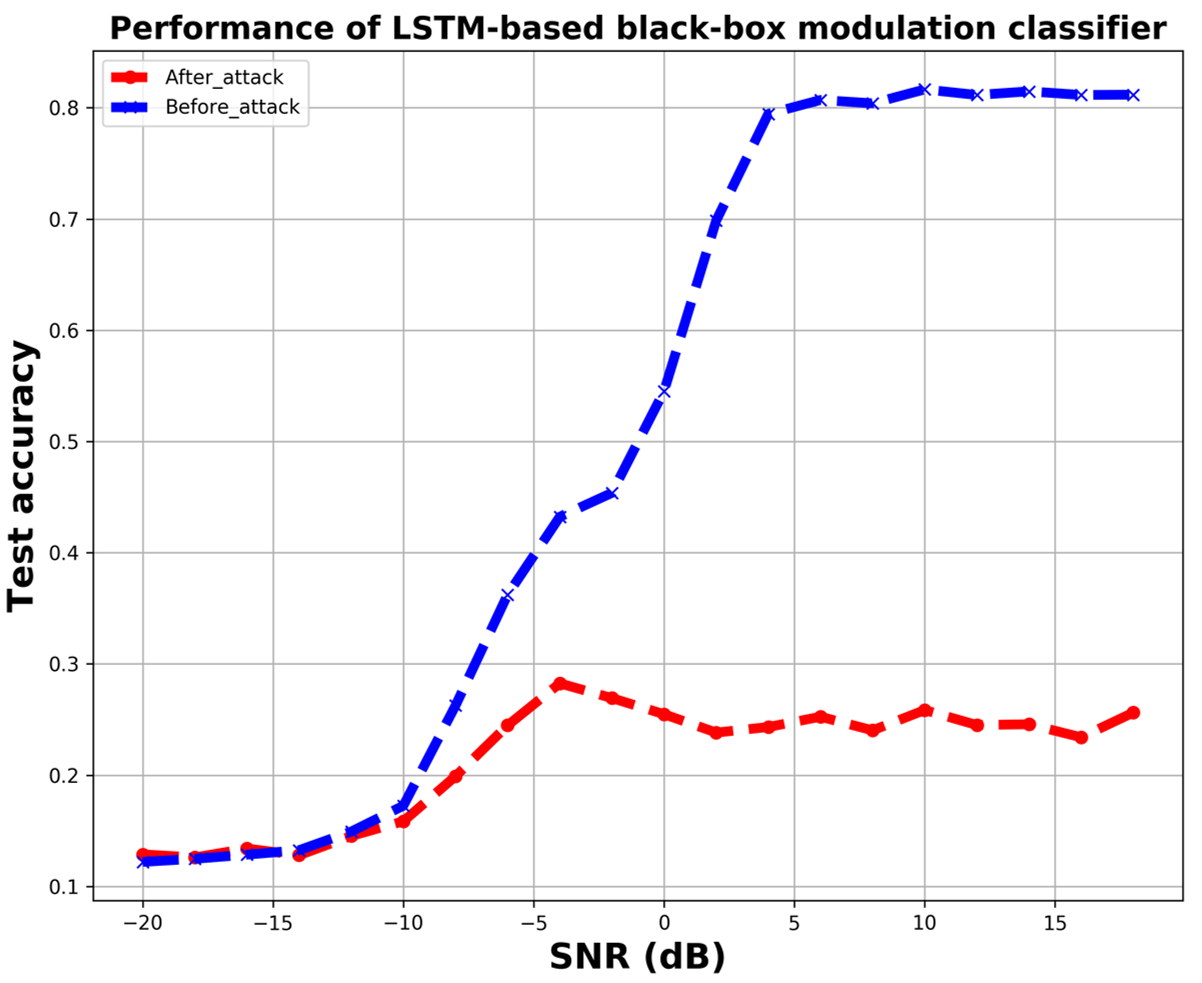}}
\caption{Performance of black-box adversarial attack on LSTM-based modulation classification.}
\end{figure}

\section{Conclusions}
In this paper, we have highlighted the lack of robustness in deep learning based modulation classification by performing a black-box adversarial attack on CNN and LSTM based modulation classifiers. We have used a surrogate deep neural network for crafting adversarial examples and then showed that adversarial examples crafted for modulation classification are transferable to other deep learning based models. We have achieved a 60\% performance drop in both CNN and LSTM based modulation classification.

%\bibliographystyle{ACM-Reference-Format}
%\bibliography{sample-base}

%\end{document}
%\endinput
%%
%% End of file `sample-sigconf.tex'.


%different levels of motion, the number of shots and encoding schemes, by using publicly available data.

%To validate our method, we carried out a comprehensive experimentation to validate the performance against different levels of motion, the number of shots and encoding schemes, by using publicly available data.

%Based on the results, we demonstrated that the performance of the proposed technique is robout to these parameters comparable to that of the state-of-the-art techniques while significantly reducing the computational time. Future plans include the extension of framework to perform end-to-end learning using generative network from motion corrupted \textit{k}-space data to artifacts free image.

% \bibliographystyle{IEEEtran}
% \bibliography{References.bib}


\end{document}